\documentclass[twocolumn,amsmath,aps]{revtex4}

\usepackage{amssymb,mathrsfs,bm,color,times}
\usepackage{graphicx,color}
\usepackage{bm}
\usepackage[hypertex]{hyperref}
\usepackage{amsmath}

\newcommand{\be}{\begin{equation}}
\newcommand{\ee}{\end{equation}}
\newcommand{\bea}{\begin{eqnarray}}
\newcommand{\eea}{\end{eqnarray}}
\newcommand{\bsube}{\begin{subequations}}
\newcommand{\esube}{\end{subequations}}

\newcommand{\Eq}[1]{Eq.\,(\ref{#1})}

\newcommand{\la}{\langle}
\newcommand{\ra}{\rangle}

\newcommand{\beq}{\begin{equation}}
\newcommand{\eeq}{\end{equation}}
\newcommand{\beqn}{\begin{eqnarray}}
\newcommand{\eeqn}{\end{eqnarray}}

\newcommand{\nl}{\nonumber \\}

\newcommand{\bsub}{\begin{subequations}}
\newcommand{\esub}{\end{subequations}}

\begin{document}

\title{Fisher information analysis on post-selection involved quantum precision measurements
using optical coherent states }

\author{Yingxin Liu}
\affiliation{Center for Joint Quantum Studies and Department of Physics,
School of Science, \\ Tianjin University, Tianjin 300072, China}

\author{Lupei Qin}
\email{qinlupei@tju.edu.cn}
\affiliation{Center for Joint Quantum Studies and Department of Physics,
School of Science, \\ Tianjin University, Tianjin 300072, China}

\author{Xin-Qi Li}
\email{xinqi.li@tju.edu.cn}
\affiliation{Center for Joint Quantum Studies and Department of Physics,
School of Science, \\ Tianjin University, Tianjin 300072, China}

\date{\today}

\begin{abstract}
{\flushleft The weak-value-amplification (WVA) }
technique has been extensively considered and debated
in the field of quantum precision measurement,
largely owing to the reduced Fisher information
caused by the low probability of successful post-selection.
In this work we show that, rather than the Gaussian meter state as typically considered,
using the optical coherent state as a meter, the WVA measurement can definitely
outperform the conventional measurement not involving the strategy of post-selection.
We also show that the post-selection procedure involved in the WVA scheme
can make a mixture of coherent states work better than a pure coherent state
with identical average photon numbers.
This is in sharp contrast to the claim proved in the absence of post-selection.
The post-selection strategy can also result in the precision of
Heisenberg (or even ``super-Heisenberg") scaling with the photon numbers,
but without using any expensive quantum resources.
The present work may stimulate further investigations
for the potential of the post-selection strategy in quantum precision measurements.
\end{abstract}


\maketitle

\section{Introduction}

{\flushleft The concept }
of quantum weak values \cite{AAV88,AV90}
has been found useful in quantum metrology.
A novel scheme termed as weak-value amplification (WVA)
has been developed for detecting tiny effects
such as the spin Hall effect of light \cite{Kwi08}
and measuring weak signals \cite{How09a,How09b,How10a,How10b,Guo13},
owing to the advantages of
improving the precision of weak measurements by post-selection
and protecting the weak measurements against technical noises
\cite{Sim10,Ste11,Nish12,Ked12,Jor14,Li20,Bru15,Bru16,How17}.

To be more specific,
for the measurement of small transverse deflections of an optical beam
\cite{How09a,How09b,How10a,How10b}, the WVA technique can lead to
experimental sensitivity beyond the detector's resolution.
In this context,
it allowed to use high power lasers with low power detectors
while maintaining the optimal signal-to-noise ratio (SNR),
and obtained the ultimate limit
in deflection measurement with a large beam radius.
Actually, theoretical analysis has pointed out that the WVA technique
can outperform conventional measurement
in the presence of detector saturation \cite{Lun17}.
This advantage of the WVA technique was recently illustrated
by achieving a precision of 6 times higher than that of
the conventional measurement \cite{ZLJ20}.
Other improvements along the line of the WVA technique include such as
strengthening the WVA via photon recycling \cite{Jor13,Sim15,Jor21}
and applying a technique called dual WVA (DWVA) which allows for
a precision metrology with sensitivity several orders of
magnitude higher than the standard approach \cite{Zen19}.
Also, it has been pointed out that the WVA technique can reduce
the technical noise in some circumstances
\cite{Kwi08,How09a,How09b,Sim10,Ste11,Nish12,Ked12,Jor14,Li20},
and can even utilize it to outperform standard measurement
by several orders of magnitude by means of the imaginary weak-value measurements
\cite{Sim10,Ste11,Ked12,Jor14,Li20}.

The WVA technique involves an essential procedure of post-selection
to the state of the measured system.
It is true that the meter state conditioned on the successful post-selection
contains more information about the parameter to be estimated
than the meter state in conventional measurement for parameter estimation.
However, since the probability of successful post-selection is low,
the practically attainable Fisher information will be greatly reduced.
It is largely this balance consideration that has caused
controversial debates in literature
\cite{Nish12,Ked12,Jor14,Li20,Tana13,FC14,Kne14,ZLJ15,Aha15}.
In Ref.\ \cite{Jor14}, using the Fisher-information (FI) metric,
it was proved that WVA technique can put all of the Fisher information
about the parameter to be estimated into a small portion of the events
and show how this fact alone gives technical advantages.
Meanwhile, there existed diverse opinions about the WVA advantages
\cite{Tana13,FC14,Kne14,ZLJ15,Aha15}.
For instance, in Ref.\ \cite{ZLJ15}, it was concluded that both the weak measurement
and post-selection cannot enhance the metrological precision.
The analysis was based on a consideration
that the Fisher information contained in the WVA technique
is only a part of the total Fisher information associated with
the post-selection measurement and the subsequent measurement on the meter state,
while the {\it reference standard} is the quantum Fisher information (QFI)
of the {\it joint entangled state} of the entire system-plus-meter
after coupling interaction (with the strength parameter to be estimated).
Obviously, such a reference standard is too stringent,
which is not the standard for comparison
with the {\it conventional} precision measurement,
such as considered in Refs.\ \cite{Ked12,Jor14,Li20} and in many others.

In precision metrology, another important criterion is
the scaling of the metrological precision with the resource (e.g. photon numbers),
aiming to excess the standard quantum limit (SQL) and to achieve the Heisenberg limit (HL).
The general viewpoint is that, in order to achieve the HL, quantum resources
such as the NOON states \cite{Dow08,Mac11,Ste14} are typically required.
However, it was found in Ref.\ \cite{ZLJ15} and further analyzed in \cite{Aha15},
that it is possible to achieve the HL
with weak measurement using the optical coherent state,
which is a classical resource, as a meter to coupled to a qubit system.
In this case, the information encoded in the distribution of post-selection process
(but {\it not} in the post-selected meter state)
can be scaled at the Heisenberg limit
with the average photon numbers of the coherent state.

For employing the optical coherent state as a meter, we become aware of
an interesting conclusion from an earlier study \cite{Lui10},
which states that for any linear coupling interaction, metrological precision
better than using the optical coherent state must come from {\it non-classical effect}.
Noting that this statement was proved in the absence of post-selection procedure,
we may thus raise a question:
can the post-selection involved in the WVA technique violate
the conclusion proved in Ref.\ \cite{Lui10}?
Of great interest is that, as we will see in this work, the answer is YES.

In this work, following the line of analysis in Refs.\ \cite{Ked12,Jor14,Li20},
where the Gaussian meter state
(i.e. the transverse spatial wavefunction of a light beam) was considered,
we further consider the probe meter as an optical coherent state
and explore its possible advantages in the WVA metrology.
This type of meter state is particularly relevant to
the optical cavity-QED or the solid-state circuit-QED set-up realization.
Also, as in Ref.\ \cite{Li20},
our analysis will go beyond the Aharonov-Albert-Vaidman (AAV) limit
by considering arbitrary measurement-coupling strengths.
By means of the Fisher-information metric, the first important result is that
using the optical coherent state as a probe meter,
which is cheap and not a ``quantum resource",
the WVA measurement can definitely outperform
the conventional measurement in the absence of the post-selection strategy.
This result is very different from the one of using the Gaussian meter state,
where the Fisher information contained in the WVA measurement can, at most,
reach the same as in the conventional measurement \cite{Jor14,Li20}.
In this context we will show that,
for the measurement of coherent state in the absence of post-selection,
we must consider quadrature measurement,
since the usual photon-number measurement cannot extract
the quantum Fisher information (QFI) encoded in the phase-shifted coherent state.
The second important result is that,
in the presence of the post-selection strategy,
the WVA measurement using a mixture of coherent states
can work better than using a single pure coherent state.
This result is of great interest when making a connection with
the claim in Ref.\ \cite{Lui10}, as mentioned above.
Since our analysis covers the range of arbitrary coupling strengths,
being somewhat surprising,
we find that it is possible to achieve a scaling behavior
with the average photon numbers even better than the HL,
when the measurement-coupling strength is properly increased to violate the AAV limit.

\section{Quantum Fisher Information Contained in the WVA Measurement}

{\flushleft Let us consider }
a two-state quantum system (qubit) with states
denoted by $|g\ra$ and $|e\ra$,
which is coupled to an optical meter prepared in coherent state $|\alpha\ra$,
while the meter state will be finally monitored by a classical apparatus
via photon number or field quadrature measurement, as schematically shown in Fig.\ 1.
In dispersive regime, the coupling interaction Hamiltonian
can be assumed as $H_{\rm int}=\chi \hat{\sigma}_z \hat{n}$,
where $\hat{\sigma}_z$ is the atomic Pauli operator
$\hat{\sigma}_z=|e\ra \la e|- |g\ra \la g|$
and $\hat{n}=\hat{a}^{\dagger}\hat{a}$ is the photon number operator
(with $\hat{a}^{\dagger}$ and $\hat{a}$ the creation and annihilation operators of a photon).   
Over some time, the coupling interaction will generate a unitary operation
$U=e^{i\lambda\hat{\sigma}_z \hat{n} }$ on the system-plus-meter joint state,
where $\lambda$ is the integrated interaction strength,
which is the right parameter to be metrologically estimated.

\begin{figure}
\includegraphics[scale=0.6]{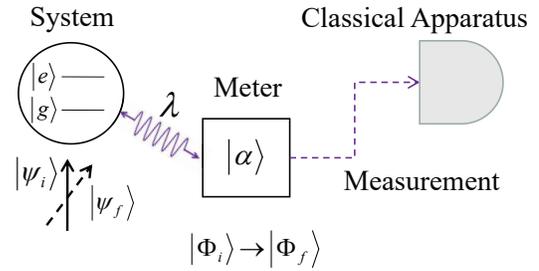}
\caption{
Sketch for precision metrology of the coupling-interaction
strength parameter ($\lambda$) between a two-state (qubit) system
and the meter of optical coherent state $|\alpha\ra\equiv|\Phi_i\ra$.
The qubit system is initially prepared in state $|\psi_i\ra$
and post-selected with $|\psi_f\ra$ after the coupling interaction.
The meter state after post-selection, $|\Phi_f\ra$,
is finally monitored by a classical apparatus
via photon number or field quadrature measurement.     }
\end{figure}

%

Assuming the system initially prepared in a superposition state characterized by
$|\psi_i\rangle={\rm cos}\frac{\theta_i}{2}|g\rangle
+{\rm sin}\frac{\theta_i}{2}e^{i\phi_i}|e\rangle$,
the coupling interaction will entangle the joint state of the system-plus-meter as
\begin{eqnarray}\label{J-state}
|\Psi_J\rangle
={\rm cos}\frac{\theta_i}{2}|g\rangle|\alpha e^{-i\lambda}\rangle
+{\rm sin}\frac{\theta_i}{2}e^{i\phi_i}|e\rangle|\alpha e^{i\lambda}\rangle  \,.
\end{eqnarray}
In the WVA strategy, a post-selection procedure is involved
by performing a projective measurement on the system,
before finally measuring the meter state for the parameter-$\lambda$ estimation.
Let us assume the post-selection with state $|\psi_f\ra$,
which has the same form of $|\psi_i\ra$
by replacing only the Bloch-vector polar angles with $\theta_f$ and $\phi_f$.
After post-selection for the system state,
the meter state reads as
\begin{eqnarray}\label{Phi-f}
&& |\tilde{\Phi}_f\rangle = \langle \psi_f|\Psi_J\rangle \nl
&&={\rm cos}\frac{\theta_i}{2}{\rm cos}\frac{\theta_f}{2}|\alpha e^{-i\lambda}\rangle
+{\rm sin}\frac{\theta_i}{2}{\rm sin}\frac{\theta_f}{2}e^{i\phi_0}
|\alpha e^{i\lambda}\rangle  \,.
\end{eqnarray}
Here we introduced $\phi_0=\phi_i-\phi_f$.
The state $|\tilde{\Phi}_f\rangle$ is not normalized.
Let us denote the normalized meter state as
$|\Phi_f\ra= |\tilde{\Phi}_f\ra / \sqrt{p_a}$,
where the normalization factor is given by
$p_a=\la \tilde{\Phi}_f|\tilde{\Phi}_f \ra$,
which is also the probability of successful post-selection,
while the subscript ``a" simply indicates the accepted case of post-selection.
This notation will be used hereafter for other quantities.   
In practice, after post-selection for the system state,
subsequent measurement to the meter state $|\Phi_f\ra$
should be done, e.g., performing the photon-number measurement
and obtaining the distribution probability $P_f(n)$.
This probability encodes the information of the parameter $\lambda$;
one can thus calculate the Fisher information $F_a$,
which characterizes the metrological precision for the parameter $\lambda$
via the Cram\'er-Rao inequality \cite{WM09}. 
In the WVA scheme, the key quantity to be compared with
the conventional approach is the {\it reduced} Fisher information $p_aF_a$ 
(to be termed as WVA-FI in this work).

If we do not specify the concrete scheme of measurement on the meter state,
we can talk about and calculate the QFI $Q_a$
encoded in the meter state $|\Phi_f\ra$, using the formula
\bea\label{QFI-a}
Q_a = 4\left[(\frac{d \la \Phi_f|}{d\lambda})(\frac{d |\Phi_f\ra}{d\lambda})
-|\la\Phi_f|(\frac{d |\Phi_f\ra}{d\lambda})|^2   \right]  \,.
\eea
Actually, the reduced QFI (or, WVA-QFI) $p_aQ_a$ represents
the maximum information attainable by the WVA measurement.
In Fig.\ 2 we display the result of $p_aQ_a$
as a function of the parameter $\lambda$.
This $\lambda$-dependence will be used later
to discuss an important conclusion,
when the WVA scheme is compared with the conventional approach.

\begin{figure}
\includegraphics[scale=0.6]{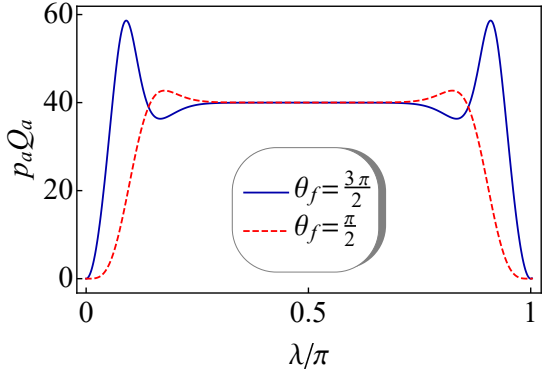}
\caption{
Dependence of the QFI encoded in the WVA scheme
on the coupling interaction strength $\lambda$.
The Bloch-vector-angles of the initial and post-selected states
of the qubit are assumed as: $\theta_i=\frac{\pi}{2}$, $\phi_0=\pi$,
and $\theta_f=\frac{3\pi}{2}$ (solid blue) and $\frac{\pi}{2}$ (dashed red). 
The average photon number $\bar{n}=|\alpha|^2=4$.  }
\end{figure}

\section{Comparison between the WVA and Conventional Measurements}

\subsection{Photon-number measurement}

{\flushleft Now let us consider }
the specific scheme of photon-number
measurement on the meter state $|\Phi_f\ra$.
The photon-number distribution probability can be predicted as
\begin{eqnarray}
&&P_f(n)={\lvert \langle n|\Phi_f\rangle \rvert}^2 \nl
&&=\frac{e^{-|\alpha|^2}}{p_a}
\left( \frac{|\alpha|^{2n}}{n!} [\, A + B\, {\rm cos}(2\lambda n+\phi_0) \, ] \right) \,.
\end{eqnarray}
Here, for simplicity of the expression, we introduced
$A={\rm cos}^{2}\frac{\theta_i}{2}{\rm cos}^2\frac{\theta_f}{2}
+{\rm sin}^{2}\frac{\theta_i}{2}{\rm sin}^2\frac{\theta_f}{2}$,
and $B= \frac{1}{2} \sin\theta_i  \sin\theta_f$.
We see clearly that this probability function encodes the information of the parameter $\lambda$,
which can be quantified by the Fisher information
\bea
F_a^{(n)} = \sum_n \frac{1}{P_f(n)} \left( \frac{dP_f(n)}{d\lambda}  \right)^2 \,.
\eea
We would like hereafter to denote
the FI associated with the photon-number measurement as $F_a^{(n)}$,
while the FI based on the $x$-quadrature measurement
is denoted as $F_a^{(x)}$
-- which will be considered below in detail.
In Fig.\ 3 (to be discussed later in detail for comparison with other schemes),
we display the numerical results of $p_aF_a^{(n)}$
as a function of the post-selection polar angle $\theta_f$.
We find that via optimal choice of $\theta_f$
the FI $p_aF_a^{(n)}$ can reach the QFI $p_aQ_a$.
Otherwise, the FI is less than the QFI, as expected.

\subsection{Field-quadrature measurement}

{\flushleft In the absence of post-selection, }
one can prove that performing photon-number measurement
on the join-state $|\Psi_J\ra$ of \Eq{J-state}
does not allow us to extract the information of $\lambda$.
That is, the FI is zero.
Therefore, if we consider the conventional approach
as considered in Refs.\ \cite{Ked12,Jor14,Li20,Lui10},
where the system is prepared in one of the basis states
(i.e. in $|g\ra$ or $|e\ra$) and no post-selection procedure is involved,
we are unable to estimate the parameter $\lambda$ via photon-number measurement,
despite that the QFI is nonzero.
For instance, consider the system prepared in $|g\ra$.
After interaction with the meter field,
the parameter $\lambda$ is encoded in the meter state as
$|\Phi_{\rm cm}\ra = |\alpha e^{-i\lambda}\ra$.
The subscript in $\Phi_{\rm cm}$ and also in the following QFI $Q_{\rm cm}$
indicates ``conventional measurement".
Using the formula of \Eq{QFI-a}, 
one can easily calculate the QFI encoded in $|\Phi_{\rm cm}\ra$ as
\bea
Q_{\rm cm}=4|\alpha|^2 = 4 \bar{n}   \,.
\eea
In this work, which is in parallel to the analysis
for the case of the Gaussian meter state,
we will use this QFI to set up the best precision
achievable by conventional measurement, when compared with the WVA scheme.

In order to extract out the information of the parameter $\lambda$
encoded in the coherent state $|\alpha e^{-i\lambda}\ra$,
rather than the photon-number measurement,
one can perform the field-quadrature measurement \cite{WM09}.
The quadrature measurement can be 
fulfilled by means of the homodyne measurement scheme, 
in which the signal field (here, the meter field) is mixed with 
a strong local oscillator of reference field with phase $\varphi$. 
Then, the homodyne readout results correspond to the eigenvalues of the quadrature operator
\bea
\hat{x}=(\hat{a}e^{-i\varphi}+\hat{a}^{\dagger}e^{i\varphi})/\sqrt{2} \,.
\eea
This measurement scheme actually defines a representation for us to construct
the wavefunction of the optical coherent state we are interested in,
i.e., the wavefunction $\alpha(x)=\la x|\alpha\ra$.
In the special case of $\varphi=0$, the $x$-quadrature operator is reduced as
$\hat{x}=(\hat{a}+\hat{a}^{\dagger})/\sqrt{2}$.
This enables us to understand the wavefunction more easily as the one of
a coherent state of the well known mechanical harmonic oscillator
in the coordinate representation.
With respect to the above $x$-quadrature operator, 
its conjugated $p$-quadrature operator is given by 
\bea
\hat{p}=-i(\hat{a}e^{-i\varphi}-\hat{a}^{\dagger}e^{i\varphi})/\sqrt{2} \,.
\eea
One can easily check that $[\hat{x},\hat{p}]=i$, i.e., the Heisenberg equation.
Now, let us consider the eigenstate equation $\hat{a}|\alpha\ra=\alpha |\alpha\ra$, 
which means that the coherent state $|\alpha\ra$
is the eigenstate of the field operator $\hat{a}$, with eigenvalue $\alpha$. 
Solving $\hat{a}$ from the two quadrature operators $\hat{x}$ and $\hat{p}$  
and substituting into the eigenstate equation, we obtain 
\bea
\frac{1}{\sqrt{2}}(\hat{x}+i\hat{p}) = \alpha e^{-i\varphi} |\alpha\ra  \,.
\eea
Then, in the $x$-representation, this equation is converted as 
\bea
\left[ \frac{d}{dx} + (x-\sqrt{2}\alpha e^{-i\varphi}) \right] \alpha(x) = 0  \,.
\eea
The solution of this differential equation reads as
\begin{equation}\label{alph-x}
\alpha(x)= (\frac{1}{\pi})^{\frac{1}{4}}
\exp \left(  \frac{1}{2}\left[ \widetilde{\alpha}^2-|\widetilde{\alpha}|^2
- (x-\sqrt{2}\widetilde{\alpha})^2  \right]   \right)   \,,
\end{equation}
where $\widetilde{\alpha}=\alpha e^{-i\varphi}$.
In this solution, the phase factor has been adjusted
in order to satisfy the property of coherent states, 
$\la\alpha_1 |\alpha_2 \ra
= \int dx \alpha^*_1(x) \alpha_2(x)
= e^{-\frac{1}{2}(|\alpha_1|^2+|\alpha_2|^2) +\alpha^*_1\alpha_2}$.     
Then, straightforwardly, the probability distribution function of the $x$-quadrature
measurement on the coherent state $|\alpha e^{-i\lambda}\rangle$ is obtained as 
\begin{equation}\label{}
P(x)={\lvert \langle x|\alpha e^{-i\lambda}\rangle \rvert}^2
= \frac{1}{\sqrt{\pi}} e^{-[x-\sqrt{2}\,
{\rm Re}(\widetilde{\alpha} e^{-i\lambda})]^2}  \,.
\end{equation}
The FI contained in this distribution function can be simply calculated as
\begin{equation}\label{FI-x}
F_{\rm cm}
=\int dx   \frac{1}{P(x)}\left(\frac{\partial P(x)}{\partial \lambda}\right)^2
= 4 {\rm Im}^2(\widetilde{\alpha} e^{-i\lambda})  \,.
\end{equation}
With no loss of generality, we assume a real $\alpha$
and obtain $F_{\rm cm}=4|\alpha|^2 \sin^2(\varphi+\lambda)$.
Then, we see that through proper choice for the local oscillator's phase $\varphi$,  
one can reach the QFI $Q_{\rm cm}=4|\alpha|^2$.

Let us now come back to the WVA scheme.
For the purpose of making a comprehensive comparison,
in addition to the photon-number measurement,
we further consider the $x$-quadrature measurement to the meter field.
After the post-selection, the meter state is a superposition of
$|\alpha e^{-i\lambda}\ra$ and $|\alpha e^{i\lambda}\ra$,
i.e., the state $|\Phi_f \ra$ given by \Eq{Phi-f}.
The probability distribution of the $x$-quadrature measurement
is given by $P_f(x)=|\la x| \Phi_f \ra |^2$, which yields
\bea
P_f(x) = | u_1\alpha_1(x) + u_2\alpha_2(x) |^2 / p_a  \,.
\eea
The two coefficients introduced here read as
$u_1=\cos\frac{\theta_i}{2}\cos\frac{\theta_f}{2}$ and
$u_2=\sin\frac{\theta_i}{2}\sin\frac{\theta_f}{2} e^{i\phi_0}$.
The two wavefunctions of the coherent states in this result
are given by $\alpha_{1,2}(x)=\la  x|\alpha e^{\mp i\lambda} \ra$,
which can be explicitly obtained using the result of \Eq{alph-x}.
Substituting $P_f(x)$ into the formula used above in \Eq{FI-x},
one can numerically compute the Fisher information
associated with the $x$-quadrature measurement,
for arbitrary post-selection options and for arbitrary interaction strengths.

\subsection{Numerical results}

{\flushleft In Fig.\ 3, }
we compare the WVA-FI associated with
the photon-number and the $x$-quadrature measurements,
$p_aF^{(n)}_a$ and $p_aF^{(x)}_a$, and also the WVA-QFI $p_aQ_a$.
First, we find that both measurement schemes can reach similar precision,
provided that the local oscillator's phase $\varphi$ in the $x$-quadrature measurement
is properly chosen. The detailed behaviors of the phase dependence are shown in Fig.\ 4.  
We emphasize that the comparison between 
the photon-number and the $x$-quadrature measurements
is possible only within the WVA scheme,
since the photon-number measurement does not make sense
in the absence of post-selection.
As expected, we find that both $p_aF^{(n)}_a$ and $p_aF^{(x)}_a$
are in general smaller than the QFI $p_aQ_a$.
But at some special post-selection angles, e.g., at $\theta_f=3\pi/2$,
both can reach the maximum value of the QFI.
In this context, we also mention that
the FI and QFI are periodic functions of $\theta_f$ with period of $2\pi$.
Here we only show the results of a single period.

\begin{figure}
\includegraphics[scale=0.5]{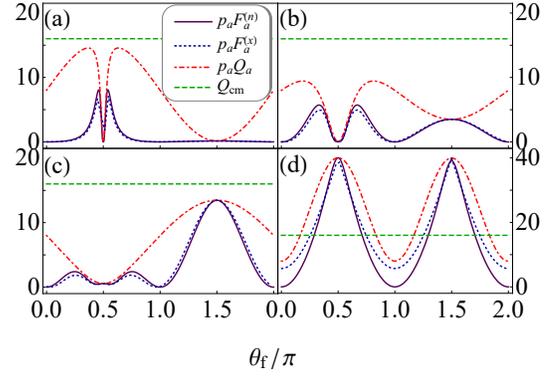}
\caption{
The WVA-FI $p_a F_a^{(n,x)}$ (solid purple and dotted blue),
associated with the photon-number and $x$-quadrature measurements,
are compared with the WVA-QFI $p_aQ_a$ (dash-dotted red)
and the QFI $Q_{\rm cm}$ of conventional metrology (dashed green).
The displayed $\theta_f$ dependence is a periodic function with period of $2\pi$.
Within a single period, the changing behaviors are shown
for a few interaction strengths:
$\lambda=0.01$ (a); 0.05 (b); 0.1 (c); and 1 (d).
Parameters used for the numerical calculations are:
$\theta_i=\frac{\pi}{2}$, $\phi_0=\pi$, and $\alpha=2$
(thus the average photon number $\bar{n}=|\alpha|^2=4$).      }
\end{figure}

\begin{figure}
\includegraphics[scale=0.7]{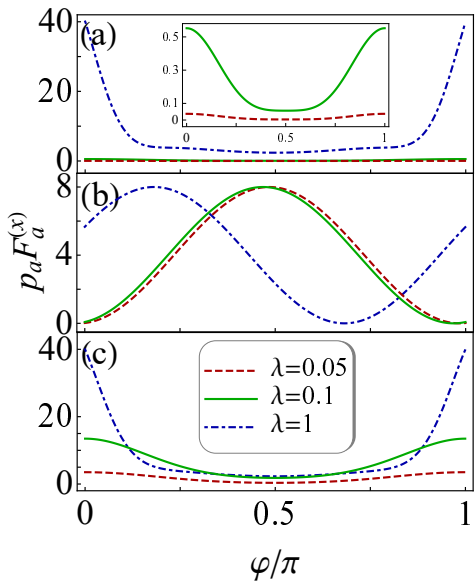}
\caption{
Dependence of the WVA-FI on the local oscillator's phase $\varphi$
in the homodyne $x$-quadrature measurement, 
after post-selection with a few angles:  
$\theta_f=\frac{\pi}{2}$ (a); $\pi$ (b); and $\frac{3\pi}{2}$ (c).
The inset in (a) is an enlarged display for the regime of small FI.
Results for interaction strengths $\lambda=0.05$, 0.1, and 1 are shown respectively
by the dashed red, solid green, and dash-dotted blue curves.
Other parameters used in the calculation are:
$\theta_i=\frac{\pi}{2}$, $\phi_0=\pi$, and $\bar{n}=|\alpha|^2=4$.     }
\end{figure}

Second, we notice that the post-selection ($\theta_f$) dependence behavior
varies drastically with the change of the interaction strength parameter $\lambda$.
Actually, this is true only in the small $\lambda$ regime.
For larger $\lambda$, i.e., the plateau area in Fig.\ 2,
the $\lambda$ dependence will become insensitive.
Very importantly, 
one finds that, with the increase of the interaction strength $\lambda$,
both the QFI $p_aQ_a$ and the FI $p_aF_a^{(n,x)}$ associated with the WVA scheme
can exceed the QFI $Q_{\rm cm}$ of the conventional metrology scheme,
as clearly observed in Fig.\ 3(d).
For broader range of parameters, this conclusion can be
anticipated as well, based on the result in Fig.\ 2.

This important observation is beyond the claim
debated for long time among the WVA community
\cite{Nish12,Ked12,Jor14,Li20,Tana13,FC14,Kne14,ZLJ15,Aha15}.
That is, it was generally believed that,
from either the perspective of Fisher information or the signal-to-noise ratio,
the {\it intrinsic} overall estimate precision cannot be enhanced by the WVA technique.
The basic reason is that, despite the enhanced signal after post-selection,
the procedure itself of post-selection
will discard a large number of measurement data
and thus result in a larger uncertainty fluctuations.
Therefore, the WVA technique was conceived of having, at least,
no theoretical advantages \cite{Tana13,FC14,Kne14,ZLJ15},
but only having some technical advantages in practice such as
not being bounded by the saturation limit of photo-detectors \cite{Lun17,ZLJ20}.

We notice that the analysis in Ref.\ \cite{Jor14} (and in many other references)
was based on a Gaussian meter state, i.e., the transverse spatial wavefunction
of the light beam in the optical realization.
Here, our analysis above is fully along the line in Ref.\ \cite{Jor14}.
The only difference is that here the meter is an optical coherent state.
Therefore, the claim that the WVA scheme cannot exceed the conventional approach
does not hold in general, 
which is {\it not} a result imposed by any fundamental principle.
The present observation in Fig.\ 3(d) is an important exception: 
it clearly shows that the WVA metrology scheme can reach a precision
much better than the conventional approach in the absence of post-selection.

\section{Mixed Coherent States}

{\flushleft In the above analysis, }
we employed the tool of Fisher information,
which is connected with the parameter estimate precision
via the well known Cram\'er-Rao inequality,
to discuss the WVA scheme
in comparison with the conventional approach (without post-selection).
In contrast to the conclusion based on the Gaussian meter state
as typically analyzed in literature \cite{Nish12,Ked12,Jor14,Li20,Tana13,FC14,Kne14},
we revealed an important metrological advantage of the optical coherent state
if the strategy of post-selection is involved.

In quantum precision metrology,
another important criterion is in concern with {\it resource consuming},
e.g., by considering the (average) photon-number ($\bar{n}=|\alpha|^2$) dependence
to be standard-quantum-limit (SQL $\propto \bar{n}$)
or Heisenberg-limit (HL $\propto \bar{n}^2$) scaling behavior.
Usually, the HL scaling can be achieved by exploiting
the quantum entanglement of photon pairs \cite{Dow08,Mac11,Ste14}.
Therefore, an interesting question is: can the HL precision
of the $\bar{n}^2$-scaling be achieved by the post-selection strategy,
without using any entangled photon pairs?

For optical coherent states, unfortunately, it was found that
the FI gained by the post-selected results of measurement
scales with $\bar{n}$ linearly \cite{ZLJ15}.
This means that only the precision of the SQL can be achieved by the WVA technique.
However, when applying the procedure of post-selection,
it was found that the FI contained in
the {\it distribution probability} of the post-selection results
is possible to achieve the HL of $\bar{n}^2$-scaling \cite{ZLJ15}.
This issue was further highlighted and analyzed in detail
by Jordan and Aharonov {\it et al} in Ref.\ \cite{Aha15}.

Remarkably, rather than using the single (pure) coherent state as the meter state,
it was found that, if using a statistical mixture of coherent states,
the HL of $\bar{n}^2$-scaling can be achieved by the WVA measurement \cite{LCF18}.
In the remainder of our present work, we would like to extend the study
from the AAV limit of extremely weak measurement, as assumed in Ref.\ \cite{LCF18},
to arbitrary strength of measurement interaction.
We will show that, with the increase of the measurement interaction strength,
the HL precision (and even better result) can be achieved in the large $\bar{n}$ regime. 
Moreover, for using classical mixed states, we will make a connection 
with the {\it contradictory statement} proved in Ref.\ \cite{Lui10},
showing that the strategy of post-selection 
may cause remarkable impact on some existing conclusions 
and make important addition to wider range of problems \cite{Lui08,Sil10}.

The statistical mixture of multiple coherent states for the optical meter
can be expressed in general as
$\rho_{\rm class}=\sum_j w_j |\alpha_j\ra  \la \alpha_j|$,
with $|\alpha_j\ra$ the individual coherent states
and $w_j$ the weighting factors.
In parallel to the formulation for the case of the pure state,
the measurement interaction between the system and the meter
is described by a unitary transformation to the initial product state,
$\rho_T = U (\rho_s\otimes\rho_{\rm class}) U^{\dagger} $,
where $U=e^{i\lambda\hat{\sigma}_z\hat{n}}$.
After post-selecting the system state with $|\psi_f\ra$, the meter state reads
$\rho_f= \la \psi_f|\rho_T|\psi_f\ra /\tilde{p}_a \equiv \tilde{\rho}_f/\tilde{p}_a$.
Now the probability of successful post-selection is given by
$\tilde{p}_a= {\rm Tr}_M [\tilde{\rho}_f]$, where ${\rm Tr}_M (\cdots)$
means the trace over a complete set of the meter states.
Then, the photon-number measurement on the meter state
gives the distribution function $P_f(n)= \la n|\rho_f|n\ra$,
and the Fisher information can be computed straightforwardly.

To be specific, let us consider a simple example of mixed state as
\bea
\rho_{\rm class}=\frac{1}{3} (|\alpha\rangle\langle\alpha|
+|{\alpha+\delta}\rangle\langle{\alpha+\delta}|
+|{\alpha-\delta}\rangle\langle{\alpha-\delta}|)
\eea
Straightforwardly, one can compute the average photon number as
$\bar{n}={\rm Tr}_M (\hat{n}\rho_{\rm class})
=|\alpha|^2+\frac{2}{3} |\delta|^2$,
and the variance of the number fluctuation as
\begin{eqnarray}\label{n-variance}
&& \left(\Delta n\right)^2 =\overline{n^2}-\left(\bar{n}\right)^2\nl
&& = \frac{2}{9} {\lvert\delta\rvert}^4
+ \frac{4}{3} \left[ |\alpha\delta|^2 + {\rm Re}(\alpha^2 \delta^{\ast 2}) \right]
+\bar{n}   \,.
\end{eqnarray}
To gain a preliminary insight, let us assume $|\delta|>>|\alpha|$.
Then, we have $\bar{n}\simeq \frac{2}{3}|\delta|^2$,
and $(\delta n)^2\simeq \frac{2}{9}|\delta|^4 \sim \bar{n}^2$.
In Ref.\ \cite{LCF18}, being restricted in the AAV limit, it was proved that
the FI is directly related with the variance of the photon numbers
of the meter state initially prepared, which simply reads
$F_a^{(n)}=4({\rm Im}\sigma^w_z)^2 (\Delta n)^2$,
where $\sigma^w_z = \la\psi_f|\sigma_z|\psi_i\ra /\la \psi_f |\psi_i\ra$
is the famous expression of the AAV weak values.
One obtains thus a metrological precision of the Heisenberg $\bar{n}^2$-scaling.

For an arbitrary strength of measurement interaction beyond the AAV limit,
after post-selection with $|\psi_f\ra$ to the system state,
the photon-number measurement on the meter state
would yield the probability distribution
\begin{eqnarray}\label{Pfn-mixed}
P_f(n) = \frac{K_n}{3\tilde{p}_a}\, [p_n(\alpha)+p_n(\alpha+\delta)+p_n(\alpha-\delta)] \,.
\end{eqnarray}
Here we have introduced the function $p_n(y)=e^{-|y|^2} |y|^{2n} / n!$
and the pre-factor
\bea
K_n = [1+\cos\theta_i\cos\theta_f
+\sin\theta_i\sin\theta_f\cos(2\lambda n+\phi_0)]/2 \,.  \nonumber
\eea
The normalization factor is given by
$\tilde{p}_a = \frac{1}{3} \sum^{3}_{j=1} p_a(\bar{n}_j)$,
where $\bar{n}_1=|\alpha|^2$ and $\bar{n}_{2,3}=|\alpha \pm\delta|^2$,
while the $p_a$ function reads as
\bea
&& p_a(\bar{n}_j) = \frac{1}{2} [\, 1+\cos\theta_i\cos\theta_f  \nl
&& ~~~~ +\sin\theta_i\sin\theta_f\,
e^{-2\bar{n}_j\sin^2 \lambda}\cos(\bar{n}_j\sin2\lambda +\phi_0) \, ] \,.   \nonumber
\eea
Using the probability distribution given by \Eq{Pfn-mixed},
the FI can be straightforwardly calculated.

\begin{figure}
\includegraphics[scale=0.75]{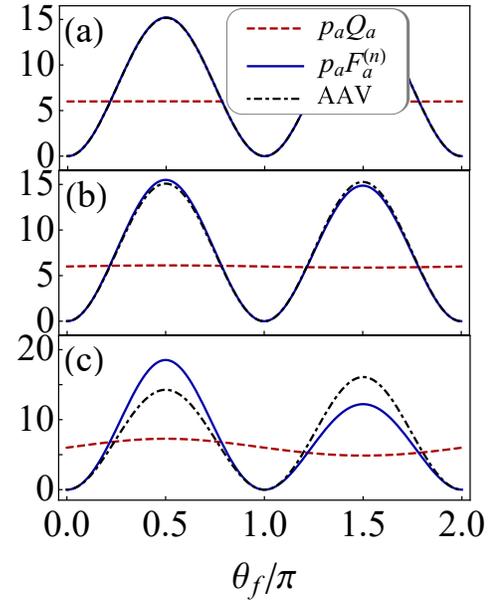}
\caption{
The WVA-FI $p_a F_a^{(n)}$
associated with photon-number measurement of mixed meter states.
The exact numerical results are compared with
the approximated ones under the AAV limit, for a few interaction strengths:
$\lambda=10^{-4}$ (a); $10^{-3}$ (b); and $10^{-2}$ (c).
The QFI $p_aQ_a$ for a pure state with the same average photon number $\bar{n}$ is displayed,
showing that the classical mixed state can work better than the quantum pure state
if employing the strategy of post-selection.
Parameters used for calculation:
$\theta_i=\frac{\pi}{2}$, $\phi_0=\frac{\pi}{2}$,
$\alpha=0.1$ and $\bar{n}=3$.      }
\end{figure}

\begin{figure}
\includegraphics[scale=0.8]{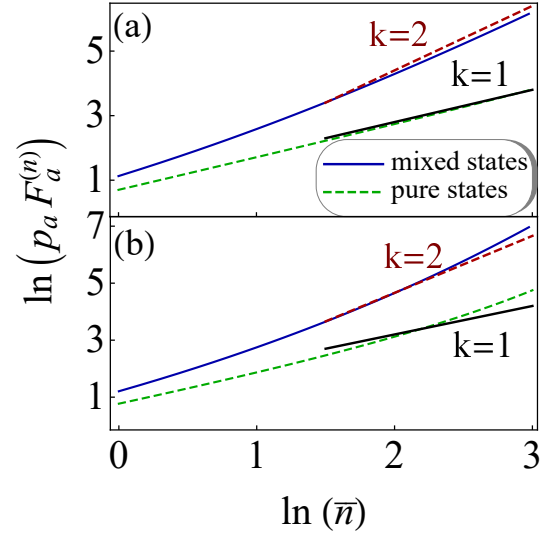}
\caption{
$\bar{n}$-scaling behavior of the WVA-FI
associated with the photon-number measurement.
Results for the interaction strengths $\lambda=10^{-3}$ and $10^{-2}$
are shown in (a) and (b), and a comparison
between the results of the mixed and pure states is plotted.
In the large $\bar{n}$ regime,
the HL scaling (and even beyond)
can be achieved by using the mixed state,
while using a pure state can largely result in the SQL scaling behavior.
Parameters used for calculation:
$\theta_i=\frac{\pi}{2}$, $\theta_f=\frac{\pi}{2}$, $\phi_0=\frac{\pi}{2}$,
and $\alpha=0.1$.     }
\end{figure}

In Fig.\ 5 we compare the WVA-FI between utilizing mixed and pure states.
Results for the former
are computed by considering the photon-number measurement,
while results for the latter are the optimal QFI $p_aQ_a$
for a pure state with identical average photon numbers $\bar{n}$.
We see clearly that the classical mixed state can work better than
the quantum pure state after applying the post-selection strategy.
To demonstrate the effect of generalization beyond the AAV limit,
we compare the exact numerical results
with the ones from the approximated expression under the AAV limit,
for a few interaction strengths.
With the increase of the strength parameter $\lambda$,
greater deviation will occur (not shown in the figure).

The important message got here is that, in the large $\bar{n}$ regime,
the mixed state can work much better than a pure state
with the same average photon numbers.
However, in Ref.\ \cite{Lui10}, it was proved that,
within the linear-coupling detection schemes (just as considered in this work),
using {\it any classical states} $\rho_{\rm class}$
(i.e. statistical mixture of coherent states)
cannot surpass the metrology precision
by using a quantum pure state $|\alpha_{\rho}\ra$ with the same $\bar{n}$.
This significant difference should originate from
the procedure of post-selection in the WVA scheme,
since in Ref.\ \cite{Lui10} only the standard measurement scheme
(i.e. without post-selection) was considered.
Therefore, we can newly claim that,
by means of involving the post-selection strategy 
as involved in the WVA scheme,
rather than what was claimed in Ref.\ \cite{Lui10},
properly designed classical mixed states
can work much better than the quantum pure states
with the same resource (i.e. the same average photon numbers).

Making a connection between the result revealed in Fig.\ 5
and the statement proved in Ref.\ \cite{Lui10} is quite constructive,
which is also relevant to a wider range of problems \cite{Lui08,Sil10},
i.e., achieving high visibility and super-resolution
{\it in the absence of quantum resources} (e.g. quantum entanglement).
It seems that the simple procedure of post-selection involved in the WVA scheme
can make an important addition to some conclusions in the field of quantum metrology,
and deeper understanding needs further investigations.

In Fig.\ 6 we reveal the $\bar{n}$-scaling behavior of the Fisher information
associated with photon-number measurement within the WVA scheme.
Results for the interaction strengths $\lambda=10^{-3}$ and $10^{-2}$
are shown in Fig.\ 6(a) and (b), and a comparison
between the results from the mixed and pure states is plotted.
In the large $\bar{n}$ regime, as indicated by the slope $k=2$ or 1 of the fitting lines,
the HL scaling (and beyond) can be achieved by using the mixed state,
while using a pure state can largely result in the SQL scaling behavior.
We emphasize that the HL scaling behavior revealed here for the mixed meter state
is also {\it induced by the post-selection procedure} in the WVA scheme,
since it was proved in Ref.\ \cite{Lui10} that, for linear coupling interaction,
any classical mixture of the coherent states
cannot work better than a single pure coherent state
(under the condition with identical average photon numbers),
while the pure state can only result in the SQL scaling
(as indicated in Fig.\ 6 by the slope $k=1$).

For the ``super-Heisenberg" scaling behavior observed in Fig.\ 6(b),
i.e., the scaling better than $k>2$,
we may provide a simple understanding as follows,
by attributing to violating the AAV limit.
Let us denote the photon-number distribution function for the mixed state as
$\bar{p}(n)= \frac{1}{J} \sum^{J}_{j=1} p_n(\alpha_j)$,
with $p_n(\alpha_j)$ the same function as defined in \Eq{Pfn-mixed}.
In the small $\lambda$ limit, the probability function of photon-numbers
measured after post-selection can be approximately expressed as
\bea
P_f(n)=\frac{K_n}{\tilde{p}}_a\,\bar{p}_(n)
\simeq \left[ 1-2\lambda{\rm Im}\sigma_z^w(n-\bar{n}) \right] \,\bar{p}_n  \,.
\eea
Then, the FI associated with this distribution function can be simply obtained as
\bea
F_a^{(n)}
= 4 ({\rm Im}\sigma_z^w )^2
\sum_n   \frac{(n-\bar{n})^2}{1-2\lambda{\rm Im}\sigma_z^w(n-\bar{n})}  \bar{p}_n   \,.
\eea
Based on this expression,
if neglecting the small $\lambda$ term in the denominator,
we obtain $F_a^{(n)}\simeq 4 ({\rm Im}\sigma_z^w)^2 \la(n-\bar{n})^2\ra$,
where the statistical average $\la\cdots\ra=\sum_n (\cdots)\bar{p}_n$ is introduced.
This result means that the photon-number scaling behavior
is determined by the initial state of the meter field,
and the HL scaling will be achieved as pointed out below \Eq{n-variance}.
However, if we keep the series expansion of $(1-z)^{-1}=1+z+z^2/2!+\cdots$
to higher orders, e.g., keeping the linear term,
we will have an additional term of contribution $\propto\la(n-\bar{n})^3\ra$,
which makes the scaling better than the HL as observed in Fig.\ 6(b).
Moreover, we have checked the $\bar{n}$ dependent behavior for
the probability $\tilde{p}_a$ of successful post-selection,
which can be approximately fitted as  
$\tilde{p}_a=d-b\bar{n}$, with the slope $b>0$.
We find that $b$ is very small for $\lambda=10^{-3}$,  
and becomes larger for $\lambda=10^{-2}$. 
This additional observation also supports the above explanation 
for the ``super-Heisenberg" scaling behavior shown in Fig.\ 6(b).
It is indeed owing to the {\it violation of the AAV limit}. 
Violation of the AAV limit also explains the scaling result
better than the SQL (with slope $k=1$),
for the pure state shown in Fig.\ 6(b).

\section{Summary and Discussion}

{\flushleft We have considered }
the WVA precision metrology
in the case of using the optical coherent state as a probe meter. 
We calculated the WVA-FI associated with two schemes
of measuring the meter's field, after post-selection to the system state.
The first scheme is the photon-number measurement as usually considered;
and the second scheme is the field-quadrature measurement 
by means of homodyne detection, which seems rarely discussed in the WVA community.
The field-quadrature measurement is necessarily required by considering the 
conventional scheme (not involving the strategy of post-selection), 
since the photon-number measurement cannot extract
the quantum Fisher information encoded in the phase-shifted coherent state.
The main important results can be briefly summarized as:
{\it (i)}
With the increase of the measurement coupling strength to violate the AAV limit,
the WVA scheme can definitely outperform the conventional approach;
{\it (ii)}
the WVA scheme can make a mixture of coherent states work better than a pure coherent state
with identical average photon numbers;
{\it (iii)}
the WVA scheme can result in the precision of
Heisenberg (or even ``super-Heisenberg") scaling with the photon numbers,
but without using any expensive quantum resources.

We notice that all the above {\it unexpected} results
are rooted in both the post-selection procedure in the WVA scheme
and the use of optical coherent states as probe meter.
For the above result {\it (i)}, when considering Gaussian meter states, 
the conclusion was very different, 
which has caused extensively controversial debates in literature.
The Gaussian meter state corresponds to 
such as the transverse spatial wavefunction in the Stern-Garlach setup,
or the one of a light beam in experiments.
In the case of optical coherent states, 
which are quite relevant to the optical cavity-QED system 
or solid-state circuit-QED architecture,
the above concluding result {\it (i)} should be much more decisive. 
For the results {\it (ii)} and {\it (iii)}, making a connection 
with the statement proved in Ref.\ \cite{Lui10} is quite constructive,
which is also relevant to a wider range of problems \cite{Lui08,Sil10},
i.e., achieving super-resolution in the absence of quantum resources.
Noting that the different conclusion in Ref.\ \cite{Lui10}
was proved in the absence of post-selection procedure,
we may expect that the simple procedure of post-selection involved in the WVA scheme
can drastically change some conclusions in the field of quantum metrology,
and deeper investigations are needed in future studies.

\vspace{2cm}
{\flushleft\it Acknowledgements.}---
This work was supported by the
National Key Research and Development Program of China
(No.\ 2017YFA0303304) and the NNSF of China (Nos.\ 11675016, 11974011 \& 61905174).


\end{document}